\documentclass[letter,twocolumn]{jpsj3}

\title{
SU(2)-SU(4) Kondo Crossover and Emergent Electric Polarization \\
in a Triangular Triple Quantum Dot
}

\author{Mikito Koga$^1$, Masashige Matsumoto$^2$, and Hiroaki Kusunose$^3$}

\inst{
$^1$Department of Physics, Faculty of Education, Shizuoka University, Shizuoka
422-8529, Japan \\
$^2$Department of Physics, Faculty of Science, Shizuoka University, Shizuoka
422-8529, Japan \\
$^3$Department of Physics, Meiji University, Kawasaki 214-8571, Japan
}

\abst{ 
We study an orbitally degenerate Kondo effect in a triangular triple quantum dot (TTQD), where the
three dots are connected vertically with a single metallic lead through electron tunneling.
Both spin and orbital degrees of freedom play an important role in the SU(4) Kondo effect.
This is demonstrated by an equilateral TTQD Kondo system at half-filling, by Wilson's
numerical renormalization group method.
We show how an emergent electric polarization of the TTQD is associated with a
crossover from SU(4) to SU(2) symmetry in the low-temperature state.
A marked sign reversal of the electric polarization is generated by the fine-tuning of Kondo
coupling with degenerate orbitals, which can be utilized to reveal orbital dynamics in the SU(4)
Kondo effect.
}

\begin{document}

\maketitle

\newcommand{\ds}{\displaystyle}
\newcommand{\bS}{{\mbox{\boldmath$S$}}}
\newcommand{\bk}{{\mbox{\boldmath$k$}}}
\newcommand{\bskp}{{\mbox{\scriptsize\boldmath $k$}}}
\newcommand{\skp}{{\mbox{\scriptsize $k$}}}
\newcommand{\bsk}{\bskp}
\newcommand{\ri}{{\rm i}}

Since Kondo's pioneering work,
\cite{Kondo64}
quantum impurities coupled to conduction electrons have
been one of the major issues of strongly correlated electron systems for over half a century.
\cite{Hewson93,Cox98,Kondo05}
Nowadays, the accumulation of knowledge about the Kondo effect in bulk systems is applied to
nanoscale or mesoscopic devices such as quantum dots (QDs) and molecular junctions, which
extends the frontiers of quantum phenomena.
\cite{Wiel02,JarilloHerrero05,Hanson07,Roch08,Parks10,Rau11}
Indeed, typical Kondo behavior is observed in the conductance between metallic leads
through a single QD, whose quantized energy levels are highly tuned by the gate
voltage.
\cite{GoldhaberGordon98,Wiel00,Potok07,Grobis08,Takada14}
Recent nanofabrication techniques have facilitated various geometric configurations of
multiple QDs.
\cite{Amaha09,Amaha13,Seo13,Folsch14}
Multicoupled QDs are considered as variants of a molecule in which spin and charge degrees
of freedom play a crucial role through strong electron correlation.
\par

Such innovation of the Kondo physics motivates us to study a triangular triple quantum dot
(TTQD) as the simplest multiple QD system with a closed loop.
\cite{Amaha09,Seo13}
When the three equivalent QDs form an equilateral triangle, doubly degenerate molecular orbitals
of TTQD play the same role as a spin.
At half-filling, the TTQD ground state is fourfold-degenerate with respect to both spin and
orbital degrees of freedom, namely, SU(4)-symmetric.
It is expected that the orbital dynamics leads to a highly symmetric SU(4) Kondo effect as a
theoretical extension of a conventional spin SU(2) Kondo effect.
\cite{Hewson93,Cox98}
Previous theoretical studies of TTQD systems have so far paid much attention to a lateral metallic contact that brings about rich Kondo physics related to three-site spin configurations of TTQD.
\cite{Kuzmenko06,Oguri07,Zitko08,Mitchell09,Vernek09,Oguri11,Koga12,Mitchell13,Oguri15}
In such a lateral geometry, it is not practical to search for the SU(4) Kondo effect since the
molecular orbital is not classified as a good quantum number.
On the other hand, it is more appropriate to investigate a different TTQD system in which each QD
is point-contacted {\it vertically} with a single lead.
In this case, the SU(4) Kondo effect is realized by the $C_3$ symmetry of TTQD.
\par

In a related context, the SU(4) Kondo physics has been frequently studied using carbon nanotube
devices whose dynamical properties come from valley degrees of freedom as well as spins,
which correspond to the clockwise or counterclockwise circular motion of electrons around a
nanotube.
\cite{JarilloHerrero05,Laird15}
In fact, the possibility of the SU(4) Kondo effect is indicated by the observation of both spin SU(2)
and valley SU(2) Kondo effects in a magnetic field.
Note that valley mixing normally lowers the SU(4) symmetry and hinders a direct
detection of the SU(4) Kondo effect.
The issue of whether the carbon nanotube can be adopted as an ideal QD device for high
controllability remains to be resolved.
\par

Keeping the above points in mind, we hereby propose a new approach to the SU(4) Kondo physics
using the TTQD system with a vertical metallic contact.
It is difficult to distinguish an SU(4) ground state from an SU(2) ground state
since both physical properties are qualitatively equivalent in the Fermi-liquid picture. 
For this reason, we focus on an emergent electric polarization due to the Kondo effect associated
with a deviation from the $C_3$ symmetry.
This is manipulated by different Kondo couplings between molecular orbitals with even and odd
parities in the TTQD geometry.
A crossover between the SU(2) and SU(4) ground states is revealed by the emergence
of an electric polarization that reflects an orbitally polarized Fermi-liquid state.
Particularly, the sign of an electric polarization depends on the orbital parity, and the SU(4)
Kondo state can be identified by its sign reversal.
\par
 
\begin{figure}
\begin{center}
\includegraphics[width=4cm,clip]{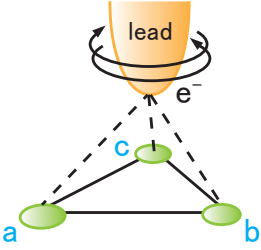}
\end{center}
\caption{(Color online)
Illustration of TTQD Kondo system.
Three QDs (green circles) are connected to a single lead vertically through electron tunneling
(broken lines).
The TTQD molecular orbitals are hybridized strongly with partial waves of lead electrons with
the $C_3$ symmetry.
Both clockwise and counterclockwise electrons are associated with degenerate $E_\pm$
orbitals relevant to the SU(4) Kondo effect.
}
\label{fig:1}
\end{figure}

We study an equilateral TTQD Kondo system in which three QDs are connected to a single
lead vertically, as shown in Fig.~\ref{fig:1}.
Since the isolated TTQD is regarded as an impurity site, the system is modeled by an impurity
Anderson Hamiltonian.
We focus on a strong Coulomb coupling case to clarify the interplay between charge, spin,
and orbital degrees of freedom associated with the Kondo effect.
The Hamiltonian examined here consists of three terms:
$H = H_{\rm dots} + H_{\rm lead} + H_{\rm l-d}$.
For the isolated TTQD, each QD is considered as an Anderson impurity and the electron motion
around the TTQD is of the hopping type:
\begin{align}
H_{\rm dots} = \sum_i (\varepsilon_{{\rm d}, i} n_i
+ U n_{i \uparrow} n_{i \downarrow})
- t \sum_{i \ne j, \sigma} d_{i \sigma}^\dagger d_{j \sigma},
\label{eqn:Hdots}
\end{align}
where $\varepsilon_{{\rm d},i}$ ($ < 0$) represents an orbital energy of the dot labelled
$i$ ($= a,b,c$), $U$ ($ > 0$) is the on-site Coulomb repulsion between two electrons occupied in
the orbital, and $t$ ($ > 0$) is an interdot electron hopping parameter.
The number operator is given by $n_i = \sum_\sigma n_{i \sigma}$
($n_{i \sigma} \equiv d_{i \sigma}^\dagger d_{i \sigma}$) for
localized electrons with spin $\sigma$ ($= \uparrow,\downarrow$) whose creation and
annihilation are expressed by $d_{i \sigma}^\dagger$ and $d_{i \sigma}$, respectively.
Here, the symmetric condition $\varepsilon_{{\rm d}, i} = - U / 2$ is used for each QD.
For the lead electrons,
\begin{align}
H_{\rm lead} = \sum_{\bsk \sigma} \varepsilon_\bsk c_{\bsk \sigma}^\dagger c_{\bsk \sigma}
\label{eqn:lead}
\end{align}
represents the kinetic energy term with $\varepsilon_\bsk$, where $c_{\bsk \sigma}^\dagger$ and
$c_{\bsk \sigma}$ are the creation and annihilation operators, respectively, for electrons with the
wavevector $\bk$ and the spin $\sigma$.
In Fig.~\ref{fig:1}, the three sites in the TTQD are coupled to the single lead through
electron tunneling matrix elements, which is considered as the hybridization between intradot
orbitals and a conduction band.
It is convenient to use the molecular orbital basis of the TTQD labelled $\tau$ ($=A, E_+ , E_-$) of a
$C_3$ irreducible representation, which is related to the three sites as
$d_{A \sigma}^\dagger =
(d_{a \sigma}^\dagger + d_{b \sigma}^\dagger + d_{c \sigma}^\dagger) / \sqrt{3}$,
$d_{E_+, \sigma}^\dagger =
(2 d_{a \sigma}^\dagger - d_{b \sigma}^\dagger - d_{c \sigma}^\dagger) / \sqrt{6}$, and
$d_{E_-, \sigma}^\dagger = (d_{b \sigma}^\dagger - d_{c \sigma}^\dagger) / \sqrt{2}$.
\cite{Janani14}
The subscripts $\pm$ respectively represent even and odd parities of doubly degenerate $E$
orbitals against the interchange between the $b$ and $c$ sites.
In this new basis, the last term in Eq.~(\ref{eqn:Hdots}) is rewritten as
\begin{align}
- t \sum_{\sigma} (2 d_{A \sigma}^\dagger d_{A \sigma}
 - d_{E_+, \sigma}^\dagger d_{E_+, \sigma} - d_{E_-, \sigma}^\dagger d_{E_-, \sigma}).
\end{align}
Since partial waves of conduction electrons with the same orbital symmetry are the most relevant
to the vertical point contact with the three QDs, we describe the hybridization with the TTQD
molecular orbitals as
\begin{align}
H_{\rm l-d} = \sum_{k \tau \sigma} (v_\tau d_{\tau \sigma}^\dagger c_{k \tau \sigma}
+ {\rm h. c.}),
\end{align}
where $v_\tau$ is assumed to be constant.
Accordingly, Eq.~(\ref{eqn:lead}) is reduced to
$H_{\rm lead} = \sum_{k \tau \sigma} \varepsilon_k c_{k \tau \sigma}^\dagger c_{k \tau \sigma}$
($k = |\bk|$) as a relevant term.
We use $\Gamma_\tau \equiv \pi \rho | v_\tau |^2$ as the strength of each point contact,
where $\rho$ is the density of states of conduction electrons at the Fermi energy.
The $E_{\pm}$ orbital degeneracy ($\Gamma_{E_+} = \Gamma_{E_-}$) is necessary for the
realization of the SU(4) Kondo effect.
For simplicity, we also assume that the fully symmetric $A$ orbital is completely localized
($\Gamma_A = 0$) since there is no contribution to the emergence of electric polarization for a
finite value of $\Gamma_A$ that does not break the $C_3$ symmetry.
In reality, $\Gamma_{E_+} \simeq \Gamma_{E_-} \gg \Gamma_A$ is a plausible
condition, by analogy with the valley degeneracy of a carbon nanotube relevant to a metallic
contact.
\cite{Laird15}
The SU(2)-SU(4) crossover in the Kondo effect can be demonstrated by introducing a deviation
from the $C_3$ symmetry of the TTQD system, which can be characterized by a finite value of
$| \Gamma_{E_+} - \Gamma_{E_-} |$.
\par

Let us begin with the low-lying energy states of the isolated TTQD at half-filling ($n_i = 1$).
For $t = 0$, the ground state is eightfold-degenerate with respect to a spin:
$S = 1/2$ (orbitally degenerate) and $S = 3/2$, where $S$ is the total spin of a TTQD state.
The interdot electron hopping ($t \ne 0$) lifts the degeneracy of $S= 1/2$ and $S = 3/2$, and it
favors the $S = 1/2$ states.
The degenerate orbital states of $S = 1/2$ are distinguished by
\begin{align}
& | \phi_{+ \uparrow} \rangle = \frac{1}{\sqrt{2}} | a \uparrow \rangle
( | b \uparrow \rangle | c \downarrow \rangle
 - | b \downarrow \rangle  | c \uparrow \rangle ),
\label{eqn:g+} \\
& | \phi_{- \uparrow} \rangle = \frac{1}{\sqrt{6}} [ | a \uparrow \rangle
( | b \uparrow \rangle | c \downarrow \rangle
 + | b \downarrow \rangle | c \uparrow \rangle )
 - 2 | a \downarrow \rangle | b \uparrow \rangle | c \uparrow \rangle ],
\label{eqn:g-}
\end{align}
where $| i \sigma \rangle$ represents a local spin state at the $i$ site in the $S_z = 1/2$ wave
functions.
The denotation $+$ ($-$) represents the even-parity spin-antisymmetric state
(odd-parity spin-symmetric state) with respect to the interchange of the $b$ and $c$ sites.
The time reversal $S_z = - 1/2$ wave functions $| \phi_{\pm \downarrow} \rangle$ are given by
the interchange of spin-up and spin-down in $| i \sigma \rangle$.
\par

In the strong Coulomb coupling limit $t / U \ll 1$, the low-energy subspace is described by
the onsite spin $1/2$ operator $\bS_i$ for the TTQD state.
Through the second-order perturbation of the hopping term in Eq.~(\ref{eqn:Hdots}), the
unperturbed $S = 1/2$ states are coupled to the excited states with double occupancy of electrons,
leading to an effective superexchange interaction.
This perturbation effect on Eqs.~(\ref{eqn:g+}) and (\ref{eqn:g-}) brings about a charge deviation
from $n_i = 1$ at each QD.
Here, we use $\delta n = ( 2 n_a - n_b - n_c ) / 3$ to measure electric polarization in the TTQD,
which is related to the following electric polarization operator expressed by the intersite spin
exchange as
\cite{Koga12,Bulaevskii08}
\begin{align}
\delta \hat{n} = 8 \left( \frac{t}{U} \right)^3 \left[ \bS_a \cdot ( \bS_b + \bS_c ) - 2 \bS_b \cdot \bS_c \right],
\label{eqn:dn1}
\end{align}
for $t / U \ll 1$.
Note that $(t / U)^3$ comes from the electron hopping around the triangular loop.
One can easily find that the charge distribution is polarized in the TTQD when the three-site spins
are not equivalent.
In the present case, $\delta n$ is generated by a finite $| \Gamma_{E+} - \Gamma_{E-} |$.
\par

Magnetic impurities are strongly coupled to conduction electrons through the
Kondo effect and the low-temperature state is described by the Kondo
singlet formation in the strong Coulomb coupling limit.
In a previous study, we reported the emergence of electric polarization in the TTQD Kondo
system for $t / U \ll 1$.
\cite{Koga12}
In that study, the target of the Kondo effect is a single local spin in the TTQD (for instance, $\bS_a$)
point-contacted by the lead, which is completely quenched through the Kondo effect with
conduction electrons at low temperatures.
According to Eq.~(\ref{eqn:dn1}), it is expected to achieve electric polarization in the TTQD as
$\delta n = -16 (t / U)^3 \langle \bS_b \cdot \bS_c \rangle$,
where $\langle \bS_b \cdot \bS_c \rangle$ is an expectation value of the ground state.
This leads to a finite $\delta n > 0$ since $\langle \bS_b \cdot \bS_c \rangle < 0$
due to the antiferromagnetic superexchange interaction.
\par

A similar mechanism also holds for the present vertical point contact with the TTQD.
The even-parity contact with $\Gamma_{E_+}$ contributes to a positive
electric polarization $\delta n > 0$ at low temperatures, while $\delta n < 0$ is expected for the
odd-parity $\Gamma_{E_-}$.
We note that the sign of $\delta n$ is related to electric polarization in the isolated TTQD given
by $\langle \phi_{\pm, \sigma} | \delta \hat{n} | \phi_{\pm, \sigma} \rangle = \pm 12 (t / U)^3$ for
$t / U \ll 1$.
\par

The model analysis of the SU(2)-SU(4) Kondo crossover is devoted entirely to investigating the
$\Gamma_{E_\pm}$ dependence of ground states and the corresponding electric polarization
$\delta n$.
The numerical calculation is carried out by Wilson's numerical renormalization group (NRG)
method using a recurrence relation at each renormalization step $N$.
\cite{Wilson75,Krishnamurthy80,Bulla08}
The original Hamiltonian is related to the $N$th NRG Hamiltonian:
$H = {\rm lim}_{N \rightarrow \infty} [D (1 + \Lambda^{-1})/2 \cdot \Lambda^{-(N-1) / 2} H_N]$,
where $\Lambda$ is a logarithmic discretization parameter of a half width $D$ of the
conduction band.
At a sufficiently large number $N$, we reach the fixed-point Hamiltonian $H_N^*$.
The corresponding energy level scheme depends on whether $N$ is even or odd
($N$ also represents the number of sites of the one-dimensional hopping-type Hamiltonian
$H_N$).
The numerical accuracy depends on the cutoff of higher-energy states, which is usually adjusted by
$\Lambda$ and another parameter $\bar{\beta} \sim 1$ related to the physical temperature
$T / D = [(1 + \Lambda^{-1})/2] \Lambda^{-(N-1) / 2} / \bar{\beta}$.
It is also useful to classify eigenstates with the orbital parity.
In the present study, we use $\Lambda = 3$, keep about 2400 lowest-lying states, and fix $U / D$ to
$0.9$.

First, let us consider the relationship between the SU(2) and SU(4) fixed points.
The former is realized for a bare value $\Gamma_{E_+} = 0$ or $\Gamma_{E_-} = 0$, while the
latter for $\Gamma_{E_+} = \Gamma_{E_-}$.
Low-lying energy levels are described by the Fermi-liquid picture.
At the SU(2) fixed point, a Kondo singlet is formed by one of the TTQD ground states
($| \phi_{+, \sigma} \rangle$ or $| \phi_{-, \sigma} \rangle$) with lead electrons.
Accordingly, one orbital ($E_+$ or $E_-$) leads to a strong coupling fixed point, and
the other is decoupled from the TTQD.
Even if $\Gamma_{E_+}$ and $\Gamma_{E_-}$ are much different in magnitude, both $E_+$ and
$E_-$ orbitals participate in the Kondo effect.
Owing to the effective exchange interaction between $E_\pm$ orbitals, an orbital
polarization of the Fermi-liquid state changes with the difference
$| \Gamma_{E_+} - \Gamma_{E_-} |$.
\par

\begin{figure}
\begin{center}
\includegraphics[width=6cm,clip]{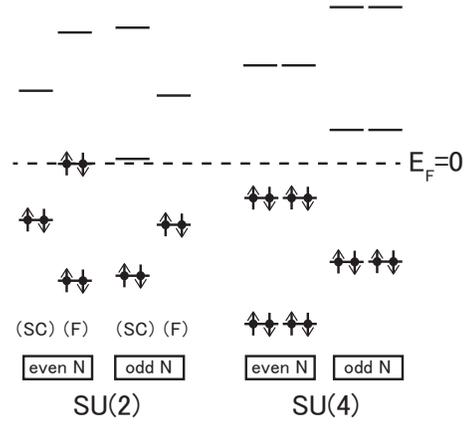}
\end{center}
\caption{
Energy spectra for ground states at the SU(2) and SU(4) fixed points of Fermi liquid in
the NRG calculation, which consist of two independent energy schemes of the $E_\pm$ orbitals.
A few lowest single-particle energy levels are shown here.
Both spin-up and spin-down electrons are occupied at each energy level below and at the Fermi
energy ($E_{\rm F} = 0$).
The strong coupling and free electron fixed points are labelled SC and F, respectively.
}
\label{fig:2}
\end{figure}
\begin{figure}
\begin{center}
\includegraphics[width=6cm,clip]{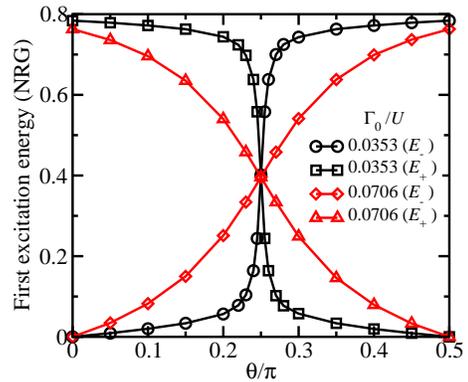}
\end{center}
\caption{(Color online)
First excitation NRG energies for the $E_\pm$ orbitals at the $\Gamma_{E_\pm}$-dependent
Fermi-liquid fixed points.
Each data (even $N$) is shown as a function of
$\theta = \arctan (\Gamma_{E_-} / \Gamma_{E_+})$,
where $\Gamma_0 \equiv (\Gamma_{E_+}^2 + \Gamma_{E_-}^2)^{1/2}$ and $t / U = 0.12$.
}
\label{fig:3}
\end{figure}
\begin{figure}
\begin{center}
\includegraphics[width=6cm,clip]{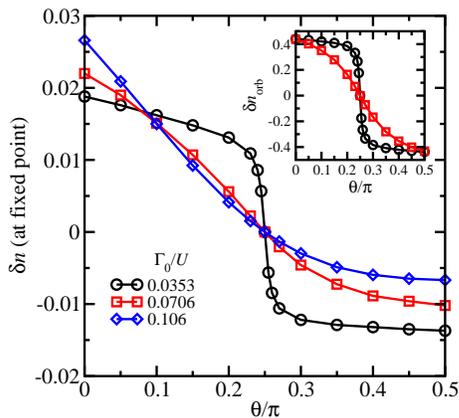}
\end{center}
\caption{(Color online)
$\Gamma_{E_\pm}$ dependence of emergent electric polarization $\delta n$ in the TTQD.
The data are shown for $\Gamma_0 / U = 0.0353, 0.0706, 0.106$ ($t / U = 0.12$).
Inset:~$\Gamma_{E \pm}$ dependence of polarization of electron occupation $\delta n_{\rm orb}$ in the $E_\pm$ orbitals for the same $\Gamma_0 / U$ in Fig.~\ref{fig:3}.
}
\label{fig:4}
\end{figure}

Quasiparticle excitations are described by discrete single-particle energy levels in the NRG.
\cite{Hewson93,Cox98,Wilson75,Krishnamurthy80,Bulla08}
The ground states for SU(2) and SU(4) are illustrated in Fig.~\ref{fig:2}, where a few lowest
one-particle energy levels are shown.
The ground state is given by filling electrons at each level from the bottom up to the Fermi
energy as the origin.
It is evident that the SU(2) fixed point is described by the combination of two independent
energy level schemes due to a strong coupling fixed point for one orbital and a free electron
fixed point for the other (labelled SC and F, respectively).
More precisely, there is a small energy shift owing to particle-hole asymmetry for SC;
for F, a single level coincides with the Fermi energy at an even $N$, leading to a
fourfold-degenerate ground state.
On the other hand, at the SU(4) fixed point, both $E_+$ and $E_-$ orbitals show the same
energy spectrum in Fig.~\ref{fig:2}.
As $| \Gamma_{E_+} - \Gamma_{E_-} |$ decreases, the ground state changes from the SU(2) fixed
point to the SU(4) fixed point continuously.
For an even $N$, each SU(2) energy level on the left (on the right) shifts upwards (downwards),
and both left and right levels merge with each other at  $| \Gamma_{E_+} - \Gamma_{E_-} | = 0$.
This also holds for an odd $N$, although the energy spectra are exchanged between SC and F.
The roles of particles and holes are interchanged between the even and odd $N$ steps.
Thus, the $E_\pm$ orbitals become degenerate at the SU(4) fixed point.
This SU(2)-SU(4) crossover can be traced by the $\Gamma_{E_\pm}$ dependence of the
first excitation energies corresponding to the $E_\pm$ orbitals in Fig.~\ref{fig:3}.
The data are shown for comparison between two values of
$\Gamma_0 \equiv (\Gamma_{E_+}^2 + \Gamma_{E_-}^2)^{1/2}$, where each energy is plotted as
a function of  $\theta = \arctan (\Gamma_{E_-} / \Gamma_{E_+})$ for a fixed $\Gamma_0$.
When $\Gamma_0$ is sufficiently small, the crossover appears with an abrupt change in the
energy level at the SU(4) fixed point ($\theta = \pi / 4$).
The s-shaped curve becomes moderate with an increase in $\Gamma_0$.
The inset of Fig.~\ref{fig:4} shows that a similar behavior appears in a polarization of
electron occupation in the $E_\pm$ orbitals, namely, an expectation value
$\delta n_{\rm orb} = \langle \sum_\sigma
(d_{E_+, \sigma}^\dagger d_{E_+, \sigma} - d_{E_-, \sigma}^\dagger d_{E_-, \sigma} ) \rangle$
in the TTQD at the $\Gamma_{E_\pm}$-dependent Fermi-liquid fixed points.
We notice that $| \delta n_{\rm orb} (\theta) | = | \delta n_{\rm orb} (\pi / 2 - \theta) |$ is satisfied in
$0 \le \theta \le \pi / 4$.
\par

It is more important to investigate the electric polarization $\delta n$ in the TTQD as a
measurable quantity, which is closely related to $\delta n_{\rm orb}$, although it does not follow the
same $\Gamma_{E_\pm}$ dependence.
In fact, $| \delta n (\theta = 0) |$ does not equal $| \delta n (\theta = \pi / 2) |$ in Fig.~\ref{fig:4},  except in the $\Gamma_0 = 0$ limit.
As in Fig.~\ref{fig:3}, $\delta n$ is plotted as a function of $\theta$
for various values of $\Gamma_0$, where $t / U$ is fixed at $0.12$.
At the even-parity SU(2) fixed point ($\theta = 0$), $| \delta n |$ increases with an
increase in $\Gamma_{E_+}$.
On the other hand, $| \delta n |$ decreases as $\Gamma_{E_-}$ increases at the odd-parity SU(2)
fixed point ($\theta = \pi / 2$).
Since a single electron is almost localized at each QD for a small $t / U$, the electric polarization
$| \delta n |$ is much smaller than the orbital polarization $| \delta n_{\rm orb} |$.
Nevertheless, $\delta n (\theta)$ also shows a sharp SU(2)-SU(4) crossover if both
$\Gamma_{E_+}$ and $\Gamma_{E_-}$ are sufficiently small. 
As $\Gamma_0$ increases, the $\delta n (\theta)$ curve becomes monotonic.
This causes difficulty in identifying an exact SU(2)-SU(4) crossover point in $\theta$.
However, one can find that $\delta n(\theta)$ approaches zero with an almost linear dependence
on $\theta$ in the $\Gamma_{E_+}$ dominant region ($0 < \theta < \pi / 4$), which implies
the SU(4) fixed point.
We also find that the magnitude $| \delta n |$ mainly depends on $t / U$.
It is approximately fitted by a power law of $t / U$ unless $t / U$ is much smaller than
$\Gamma_0 / U$ (the exponent is close to 3 at $\Gamma_0 \rightarrow 0$).
\par

\begin{figure}
\begin{center}
\includegraphics[width=6cm,clip]{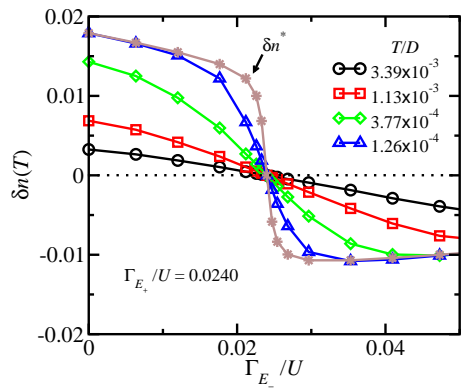}
\end{center}
\caption{(Color online)
Temperature dependence of $\delta n$ as a function of
$\Gamma_{E_-} / U$ for $\Gamma_{E_+} / U = 0.0240$ and $t / U = 0.12$.
Here, $\delta n^*$ represents an electric polarization at the $\Gamma_{E_\pm}$-dependent
fixed points.
}
\label{fig:5}
\end{figure}
In Fig.~\ref{fig:5}, we show how $\delta n$ is induced by the Kondo effect with the decrease in
temperature.
Each data point is plotted as a function of $\Gamma_{E_-} / U$, where
$\Gamma_{E_+} / U = 0.0240$
is chosen as a small constant value.
Despite a large $| \Gamma_{E_+} - \Gamma_{E_-} |$ difference, $\delta n (T)$ is close to zero at
high temperatures, indicating that the TTQD state $| \phi_{\pm, \sigma} \rangle$ maintains the
$E_\pm$ orbital degeneracy.
At low temperatures, the emergence of $\delta n$ is accompanied by the complete screening of
the TTQD spin.
At sufficiently low temperatures, $\delta n$ reaches a fixed point value $\delta n^*$.
Note that $\delta n (T) = 0$ is maintained at all temperatures for $\Gamma_{E_-} = \Gamma_{E_+}$
owing to the SU(4) symmetry.
In the vicinity of the SU(4)-symmetric point, $| \delta n |$ shows an abrupt increase with a
decrease in temperature.
From the viewpoint of application, the SU(2)-SU(4) Kondo crossover is useful for the manipulation
of the sign  reversal of the emergent electric polarization.
\par

In summary, we show that the TTQD device is an attractive tool for detecting orbital dynamics in the
SU(2)-SU(4) Kondo crossover.
The emergence of the electric polarization $\delta n$ is controlled by the difference in the Kondo
coupling strength between the $E_\pm$ orbitals that brings about a deviation from the $C_3$
symmetry of TTQD.
The SU(2)-SU(4) Kondo crossover is confirmed by a marked change in $\delta n$, which becomes
more prominent with a decrease in the point-contact strength $\Gamma_{E_\pm} $. 
The recent nanofabrication technique has achieved the confinement of a few electrons in a TTQD
system.
\cite{Amaha09,Seo13}
Electron tunneling through a point contact can be fine-tuned by adjusting the gate voltage of each
QD.
\cite{Kanai10}
Thus, the manipulation of an electric polarization on the nanoscale could be promising for a new
application of the Kondo effect.
\par

\acknowledgment
This work was supported by JSPS KAKENHI Grant Numbers 25400322, 26400332, 15K05176,
and 15H05885 (J-Physics).



\begin{thebibliography}{99}

\bibitem{Kondo64}
  J. Kondo,
  Prog. Theor. Phys. {\bf 32}, 37 (1964).

\bibitem{Hewson93}
  A. C. Hewson,
  {\it The Kondo Problem to Heavy Fermions} (Cambridge University Press, Cambridge, U.K., 1993).

\bibitem{Cox98}
  D. L. Cox and A. Zawadowski,
  Adv. Phys. {\bf 47}, 599 (1998).

\bibitem{Kondo05}
  See review articles in J. Phys. Soc. Jpn. {\bf 74} (2005) (No. 1, Special Topics).

\bibitem{Wiel02}
  W. G. van der Wiel, S. De Franceschi, J. M. Elzerman, T. Fujisawa, S. Tarucha, and
  L. P. Kouwenhoven,
  Rev. Mod. Phys. {\bf 75}, 1 (2002).
  
\bibitem{JarilloHerrero05}
  P. Jarillo-Herrero, J. Kong, H. S. J. van der Zant, C. Dekker, L. P. Kouwenhoven, and
  S. De Franceschi,
  Nature {\bf 434}, 484 (2005).
  
\bibitem{Hanson07}
  R. Hanson, L. P. Kouwenhoven, J. R. Petta, S. Tarucha, and L. M. K. Vandersypen,
  Rev. Mod. Phys. {\bf 79}, 1217 (2007).

\bibitem{Roch08}
  N. Roch, S. Florens, V. Bouchiat, W. Wernsdorfer, and F. Balestro,
  Nature {\bf 453}, 633 (2008).

\bibitem{Parks10}
  J. J. Parks, A. R. Champagne, T. A. Costi, W. W. Shum, A. N. Pasupathy, E. Neuscamman,
  S. Flores-Torres, P. S. Cornaglia, A. A. Aligia, C. A. Balseiro, G. K.-L. Chan,
  H. D. Abru{\~{n}}a, and D. C. Ralph,
  Science {\bf 328}, 1370 (2010).

\bibitem{Rau11}
  I. G. Rau, S. Amasha, Y. Oreg, and D. Goldhaber-Gordon,
  in {\it Understanding Quantum Phase Transitions}, ed. L. D. Carr
  (CRC Press, Boca Raton, FL, 2010) Chap.~14.

\bibitem{GoldhaberGordon98}
  D. Goldhaber-Gordon, H. Shtrikman, D. Mahalu, D. Abusch-Magder, U. Meirav, and
  M. A. Kastner,
  Nature {\bf 391}, 156 (1998).

\bibitem{Wiel00}
  W. G. van der Wiel, S. De Franceschi, T. Fujisawa, J. M. Elzerman, S. Tarucha, and
  L. P. Kouwenhoven,
  Science {\bf 289}, 2105 (2000).
  
\bibitem{Potok07}
  R. M. Potok, I. G. Rau, H. Shtrikman, Y. Oreg, and D. Goldhaber-Gordon,
  Nature {\bf 446}, 167 (2007).
  
\bibitem{Grobis08}
  M. Grobis, I. G. Rau, R. M. Potok, H. Shtrikman, and D. Goldhaber-Gordon,
  Phys. Rev. Lett. {\bf 100}, 246601 (2008).
  
\bibitem{Takada14}
  S. Takada, C. B{\"{a}}uerle, M. Yamamoto, K. Watanabe, S. Hermelin, T. Meunier, A. Alex,
  A. Weichselbaum, J. von Delft, A. Ludwig, A. D. Wieck, and S. Tarucha,
  Phys. Rev. Lett. {\bf 113}, 126601 (2014).

\bibitem{Amaha09}
  S. Amaha, T. Hatano, T. Kubo, S. Teraoka, Y. Tokura, S. Tarucha, and D. G. Austing,
  Appl. Phys. Lett. {\bf 94}, 092103 (2009).

\bibitem{Amaha13}
  S. Amaha, W. Izumida, T. Hatano, S. Teraoka, S. Tarucha, J. A. Gupta, and D. G. Austing,
  Phys. Rev. Lett. {\bf 110}, 016803 (2013).
  
\bibitem{Seo13}
  M. Seo, H. K. Choi, S.-Y. Lee, N. Kim, Y. Chung, H.-S. Sim, V. Umansky, and D. Mahalu,
  Phys. Rev. Lett. {\bf 110}, 046803 (2013).

\bibitem{Folsch14}
  S. F{\"{o}}lsch, J. Mart{\'{i}}nez-Blanco, J. Yang, K. Kanisawa, and S. C. Erwin,
  Nat. Nanotechnol. {\bf 9}, 505 (2014).

\bibitem{Kuzmenko06}
  T. Kuzmenko, K. Kikoin, and Y. Avishai,
  Phys. Rev. Lett. {\bf 96}, 046601 (2006).
  
\bibitem{Oguri07}
  A. Oguri, Y. Nisikawa, Y. Tanaka, and T. Numata,
  J. Magn. Magn. Mater. {\bf 310}, 1139 (2007).

\bibitem{Zitko08}
  R. {\v{Z}}itko and J. Bon{\v{c}}a,
  Phys. Rev. B {\bf 77}, 245112 (2008).

\bibitem{Mitchell09}
  A. K. Mitchell, T. F. Jarrold, and D. E. Logan,
  Phys. Rev. B {\bf 79}, 085124 (2009).
  
\bibitem{Vernek09}
  E. Vernek, C. A. B{\"{u}}sser, G. B. Martins, E. V. Anda, N. Sandler, and S. E. Ulloa,
  Phys. Rev. B {\bf 80}, 035119 (2009).

\bibitem{Oguri11}
  A. Oguri, S. Amaha, Y. Nishikawa, T. Numata, M. Shimamoto, A. C. Hewson, and S. Tarucha,
  Phys. Rev. B {\bf 83}, 205304 (2011).

\bibitem{Koga12}
  M. Koga, M. Matsumoto, and H. Kusunose,
  J. Phys. Soc. Jpn. {\bf 81}, 123703 (2012).

\bibitem{Mitchell13}
  A. K. Mitchell, T. F. Jarrold, M. R. Galpin, and D. E. Logan,
  J. Phys. Chem. B {\bf 117}, 12777 (2013).

\bibitem{Oguri15}
  A. Oguri, I. Sato, M. Shimamoto, and Y. Tanaka,
  J. Phys.: Conf. Ser. {\bf 592}, 012143 (2015).
  
\bibitem{Laird15}
  E. A. Laird, F. Kuemmeth, G. A. Steele, K. Grove-Rasmussen, J. Nyg{\r{a}}rd, K. Flensberg, and
  L. P. Kouwenhoven,
  Rev. Mod. Phys. {\bf 87}, 703 (2015).

\bibitem{Janani14}
  C. Janani, J. Merino, I. P. McCulloch, and B. J. Powell,
  Phys. Rev. Lett. {\bf 113}, 267204 (2014).

\bibitem{Bulaevskii08}
  L. N. Bulaevskii, C. D. Batista, M. V. Mostovoy, and D. I. Khomskii,
  Phys. Rev. B {\bf 78}, 024402 (2008).

\bibitem{Wilson75}
  K. G. Wilson,
  Rev. Mod. Phys. {\bf 47}, 773 (1975).

\bibitem{Krishnamurthy80}
  H. R. Krishna-murthy, J. W. Wilkins, and K. G. Wilson,
  Phys. Rev. B {\bf 21}, 1003 (1980);
  H. R. Krishna-murthy, J. W. Wilkins, and K. G. Wilson,
  Phys. Rev. B {\bf 21}, 1044 (1980).

\bibitem{Bulla08}
  R. Bulla, T. A. Costi, and T. Pruschke,
  Rev. Mod. Phys. {\bf 80}, 395 (2008).

\bibitem{Kanai10}
  Y. Kanai, R. S. Deacon, A. Oiwa, K. Yoshida, K. Shibata, K. Hirakawa, and S. Tarucha,
  Phys. Rev. B {\bf 82}, 054512 (2010).
  
\end{thebibliography}
\end{document}